\documentclass[usenatbib]{mn2e}
\usepackage{color} 

\usepackage{graphics}
\usepackage{epsfig}
\usepackage{natbib}

\begin{document}

\title[Morphology and kinematics of neutral hydrogen]
{The morphology and kinematics of neutral hydrogen in the vicinity of z=0 galaxies with Milky Way masses
-- a study with the Illustris simulation} 

\author [G.Kauffmann et al.] {Guinevere Kauffmann$^1$\thanks{E-mail: gamk@mpa-garching.mpg.de},
Sanchayeeta Borthakur$^2$, Dylan Nelson$^1$\\  
$^1$Max-Planck Institut f\"{u}r Astrophysik, 85741 Garching, Germany\\
$^2$Department of Physics \& Astronomy, Johns Hopkins University,
Baltimore, MD, 21218, USA}

\maketitle

\begin{abstract} 
We analyze the properties of the circumgalactic gas (CGM) around 120 galaxies with stellar and dark
matter halo masses similar to that of the Milky Way. We focus on the 
morphology and kinematics of the neutral hydrogen and how this depends on    
$f_g$, the ratio of gas-to-stellar mass within the optical radius.      
In gas-rich galaxies with $f_g > 0.1$, gas
temperatures rise slowly from center of the halo out to the virial radius and  
average neutral gas column densities remain  above 
$10^{19}$ atoms cm$^{-2}$ 
out to radii of 50-70 kpc. In gas-poor
galaxies with $f_g < 0.1$, gas temperatures rise quickly outside the edge of the
disk to  $\sim 10^6$ K,
and then remain fixed out to radii of 100 kpc.
The column density of neutral gas quickly drops  below $10^{19}$ atoms cm$^{-2}$ at
radii of 10 kpc.     
Neutral gas distributions are also more asymmetric  in 
gas-poor galaxies. Most of the differences between gas-poor and gas-rich galaxies 
in the Illustris simulation can be attributed to the effects of
``radio-mode" AGN feedback. 
In the Illustris simulation, the circumgalactic gas is found to rotate coherently about the center of the
galaxy with a maximum rotational velocity of around 200 km/s.  In gas-rich
galaxies, the average coherence length of the rotating gas is 40 kpc,
compared to 10 kpc in gas-poor galaxies. In the very most gas-rich systems,
the CGM can rotate coherently  over scales of 70-100 kpc.
We discuss our results in the context of recent observations of the CGM
in low mass galaxies via UV absorption-line spectroscopy and deep 21cm
observations of edge-on spiral galaxies,
\end{abstract}


\section{Introduction}

The accretion of gas onto galaxies from the external environment remains
poorly understood.  The main obstacle to making progress through direct
observations is that data at a single frequency provides information on a
very restricted range of gas densities and temperatures.  In particular,
the presence of a background radiation field of ultraviolet radiation
produced by quasars and star-forming galaxies results in  much of the
hydrogen in the halos of galaxies being  ionized and inaccessible to radio
observations of 21cm emission from neutral hydrogen.  The ionized gas
component is most easily probed by absorption lines in the spectra
of background quasars. The main limitation is  that each quasar
sightline provides a purely 1-dimensional view of the intervening gas and
it is only possible to reconstruct the density profiles and the structure
of the circumgalactic gas statistically by combining large numbers of
quasar spectra that pass through the halos of many different galaxies (see
Rudie et al 2012; Tumlinson et al 2013; Werk et al 2015; Borthakur et al
2015 for recent examples of studies of this kind).

In principle, cosmological hydrodynamical simulations provide a means of
{\em predicting} how gas accretes onto galaxies. Early work emphasized a
dichotomy between so-called ``cold accretion'' directed along filaments
in dark matter halos with masses less than $\sim
10^{11.4} M_{\odot}$, and a quasi-spherical ``hot mode'' of accretion in
more massive halos (Keres et al 2005). These results were later shown to be
sensitive both to the numerical implementation used to solve the equations
of hydrodynamics (Nelson et al. 2013), and to the implementation of
feedback from supernova and active galactic nuclei (Nelson et al. 2015b).
However, early generations of cosmological hydrodynamical simulations also generally
failed to match basic observational quantities, including the stellar mass-to-halo mass relation
for galaxies (Guo et al 2010) and the size distribution function of galaxies at
fixed mass (Weil, Eke \& Efstathiou 1998).  Most simulations also failed to reproduce  the relative
fractions of star-forming and non-star-forming objects in the real
Universe (e.g. Cattaneo et al 2007), and were thus unlikely to 
provide accurate predictions for the
properties of the gas surrounding galaxies of different masses and types.

In the last year or two, the situation has changed quite dramatically.  The
Illustris (Vogelsberger et al 2014) and Eagle (Schaye et al 2015)
simulation projects have published results from large (100 Mpc$^3$) volume
simulations with gas/dark matter mass resolutions of a few $\times 10^6
M_{\odot}$, which contain many hundreds of galaxies of Milky Way mass that
are simulated with spatial resolutions of 0.7-1 kpc.  The physical inputs,
e.g. the supernovae feedback prescriptions, are tuned to reproduce
observed quantities. In the Illustris simulations, for example, the model
is tuned to match the present-day ratio of stars to dark
matter (DM) for galaxies of all masses, and the total amount of star
formation in the universe as a function of time. Other simulation projects,
e.g. NIHAO (Wang et al 2015) and FIRE (Hopkins et al 2013), have
concentrated on generating suites of cosmological zoom-in hydrodynamical
simulations of haloes of different masses.  Once again, the aim has been to
reproduce the stellar mass versus halo mass and the star formation rate versus stellar
mass relations using the same treatment of star formation and stellar
feedback for every object.

These simulations are now starting to form the basis for studies of gas
around galaxies as a function of cosmic epoch. One approach that has been
popular is to use the simulations to generate mock quasar absorption line
spectra for direct comparison with observations. This is done by averaging
gas physical properties including the  density, temperature, metallicity
and velocity in physical coordinates along a sightline through the
simulation box, and then using  look-up tables calculated with a
photo-ionization code to find the ionization fraction for a number of
different ionic species (see for example Ford et al 2013).  This
methodology has now been implemented on the latest generation of
simulations that reproduce stellar mass to dark matter mass ratios, and the
mock spectra have been used to make comparisons with hydrogen and metal
line absorption around both low-redshift (Suresh et al 2015a; Oppenheimer et al 2016) and high
redshift (Shen et al 2013; Bird et al 2014; Rahmati et al 2014; Faucher-Gigu{\`e}re et al 2015;
Suresh et al 2015b) galaxies. Most of these analyses have focused either on
comparisons of the radial dependence of Ly$\alpha$ and metal absorption
line equivalent widths, or on the {\em covering fraction} of neutral
hydrogen and metals, defined as the {\em fraction of sightlines} where
intervening neutral hydrogen or metals are detected in the spectrum.

At high redshifts, some of the simulations have had
trouble explaining the large observed covering fractions of
neutral hydrogen at impact parameters comparable to the
virial radius of the galaxy (Faucher-Gigu\`ere et al 2015;
Suresh et al 2015a), while others apparently produce better
agreement with the data (Rahmati et al 2015).
Also of interest are trends in CGM
properties as a function of galaxy mass and type.  In very recent work,
Suresh et al (2015b) examine whether the simulations are able to explain
trends in OVI properties as functions of galaxy mass and star formation rate
at low redshifts using the Illustris simulations. 
The mass, temperature and metallicity of the CGM were found to increase
with stellar mass, driving an increase in the OVI column
density profile, which is not seen in observations. More recently,
Oppenheimer et al (2016) have performed a similar
study with the EAGLE simulations and find that gas temperatures
are high enough in group haloes to suppress the
OVI column density, leading to better agreement with the data.
Taken together, these studies illustrate the diagnostic
power of Ly$\alpha$ and  metal absorption line analyses around
galaxies.

In this paper, we take a somewhat different approach and concentrate on the
predicted temperature and density structure, morphology and kinematics of
the circumgalactic gas in the Illustris simulations. The morphology of the
gas has been probed to a limited extent in  absorption line studies by
studying trends in Ly$\alpha$ equivalent width as a function of orientation
with respect to the stellar disk (Borthakur et al 2015), while the velocity
spread of the gas measured from the distribution of velocity offsets of the
centroid of the primary Ly$\alpha$ component from the systemic velocity has
been proposed as a test of neutral gas kinematics (Rudie et al 2012; Liang \& Chen
2014; Borthakur et al 2015).  These probes are crude, because they are
derived by combining sightlines around multiple galaxies and cannot, for
example, probe whether the CGM is co-rotating with the stellar disk or not.
As we will discuss,  direct imaging of low column density neutral and
ionized gas (e.g. Kamphuis et al 2013; Zschaechner, Rand \& Walterbos 2015)
may provide additional direct tests of simulation predictions.

Our paper is organized as follows. In section 2, we briefly summarize
pertinent details of the Illustris simulations and describe our analysis
strategy.  In section 3, we present our results and in section 4, we
discuss these results in light of recent and future observational
constraints.

\section{Simulations and Analysis}

\subsection {Simulation details}

The Illustris simulations are run using the AREPO code (Springel 2010),
a finite volume scheme implemented on a moving mesh, which follows
the gas flow in a quasi-Langragian fashion. As a result, AREPO retains the
principal strengths of SPH including its adaptivity to a large dynamic
range in spatial scales, Galilean  invariance,  and  an  accurate  and
efficient gravity solver. It also gains the strengths of finite volume
codes, including the improved treatment of fluid instabilities, weak shocks
which can be missed in SPH, phase interfaces and shearing flows.

We make use of data from the Illustris-1 simulations, whose properties are
summarized in Table 1 of Nelson et al (2015a). This is the largest and
highest resolution of the publicly released simulations with a volume of
106.5 Mpc$^3$, dark matter and gas particle masses of $6.3 \times 10^6
M_{\odot}$ and $1.6 \times 10^6 M_{\odot}$, and dark matter and gas
gravitational softening lengths of 1.4 and 0.7 kpc. The reader is referred
to Vogelsberger et al. (2013) for details on the behavior, implementation,
parameter selection, and validation of the physical models implemented in
these simulations.

Of particular interest for  results on the distribution and kinematics of
neutral hydrogen in galaxy halos presented in this paper is the
implementation of primordial and metal-line radiative cooling in the
presence of a redshift-dependent, spatially uniform, ionizing UV background
field, with self-shielding corrections, and the  implementation of
galactic-scale outflows from supernovae and feedback from AGN in both
quasar and radio (bubble) modes. In brief:

{\bf Gas cooling.} The code includes  a self-consistent calculation of the
primordial cooling following Katz et al. (1996), on top of which  the
cooling contribution of metals is added in a simplified way by assuming
that the cooling rate scales linearly with the total metallicity of the
gas. Photoionization rates from quasars and star-forming galaxies, which
affect abundances and inject energy into the gas, are calculated based on
the models  of Faucher-Gigu{\`e}re et al. (2009).
In practice, the cooling rates
are evaluated from tabulated CLOUDY models which include
as input gas temperature, density, metallicity, and the
UV-background.

This treatment of gas cooling and heating is valid under the assumption
that gas is optically thin to the ambient UV radiation and will break down
at gas densities greater than $10^{-3}$ cm$^{-3}$, above which the gas
absorbs radiation and the attenuated radiation field affects the balance
between cooling and heating compared to the optically thin case.  In
Illustris, a ``self-shielding'' correction whereby the ionization and
heating rates entering into the primordial network calculation  are
suppressed.
The supression factors are based on the results of full radiative
transfer calculations carried out by Rahmati et al (2013).

{\bf Supernova feedback.} The 
implementation of supernova feedback is very similar to the scheme
described in Springel \& Hernquist (2003) and operates by probabilistically
turning a star-forming ISM gas cell into a wind particle.
The probability depends on the instantaneous SFR of each
gas cell, which is derived from the local gas density.
This particle
is then allowed to travel, decoupled from hydrodynamical forces, until it
reaches a certain density threshold or a maximum travel time has elapsed,
whereupon the particle disappears and its  mass, momentum, thermal energy
and all tracked metals are deposited into the gas cell in which it is
located.  The wind velocity and the mass loading of the particles are free
parameters and in the fiducial Illustris model, the wind particle is kicked
perpendicular to the angular momentum vector of the parent subhalo of the
galaxy (corresponding to a bipolar wind).  

{\bf AGN feedback} Two modes of AGN feedback are implemented. When the
central black hole accretes at rates higher than about 5\% of the Eddington
rate, the so-called ``quasar mode'' (Springel et al 2005) is implemented by
assuming that a fraction of the radiative energy released by the accreted
gas couples thermally to nearby gas within a radius that contains some
fixed amount of mass.  When the quasar mode is switched on, Illustris also
accounts for the effect of the AGN radiation field on the cooling of the
gas.

This paper is concerned with CGM properties in low redshift galaxies, where
the vast majority of black holes accrete at rates of only a few percent of
the Eddington rate.  For these low-activity states of the black hole, a
form of mechanical radio-mode AGN feedback following Sijacki et al. (2007)
is implemented.  Bubbles of hot gas with radius 50 kpc, total energy
$10^{60}$ erg and volume density $10^4 M_{\odot}$ kpc$^{-3}$ are placed into
the halo at distances of $\sim$100 kpc from their centers.  The radio-mode
feedback efficiency provided by the bubbles is assumed to be a fixed
fraction of the rest mass energy of the gas accreted by the black hole.

\subsection {Analysis} The analysis in this paper focuses on galaxies with
masses comparable to those of the Milky Way, i.e. galaxies with stellar
masses (calculated within twice the stellar half mass radius) in the range
$6-8 \times 10^{10} M_{\odot}$ residing in subhalos with dark matter masses
in the range $1-2 \times 10^{12} M_{\odot}$. The CGM of these galaxies is
well-resolved in Illustris-1, with around  160,000 dark matter particles,
15,000-30,000 gas cells and 80,000-100,000 star particles contained within
the virial radius of the subhalo. 
Out of the 253 galaxies in this stellar
mass range, we select 120 galaxies at z=0 (snapshot number
135) such that they equally populate 4 bins in gas mass
fraction, $f_g=
M_{gas}/M_{stars}$, where $M_{gas}$ is again evaluated within twice the
stellar half mass radius. The four bins in $f_g$ are 0.01-0.03, 0.03-0.1,
0.1-0.3 and 0.3-1.
This is the maximal
sample such that the equal number per $f_g$ bin criterion is
satisfied. Note that we do not implement any prescription
for the fraction of gas in molecular form, so the parameter $f_g$
should be a close approximation to the total cold gas mass
fraction interior to the optical radius of the galaxy.

In Figure 1, we show distributions of the total cold
gas mass fraction of galaxies in three different bins of stellar
mass. Solid black histograms show results from Illustris,
while red histograms show results from the COLD
GASS survey (Saintonge et al 2011), which obtained atomic
and molecular gas masses for a sample of 350 representative
nearby galaxies with stellar masses in the range
$10^{10}-10^{11.5} M_{\odot}$. The central panel correpsonds to galaxies
with stellar masses comparable to that of the Milky Way.
We see that the dynamic range in gas fraction spanned by
Illustris is larger than in the data; there are too few very
gas-poor galaxies of Milky Way mass and too many very
gas-rich galaxies. The median gas fractions of the simulated
and observed data sets agree quite well, however, and are
close to the observed value for the Milky Way ($f_g \sim  0.1$).

\begin{figure*}
\includegraphics[width=160mm]{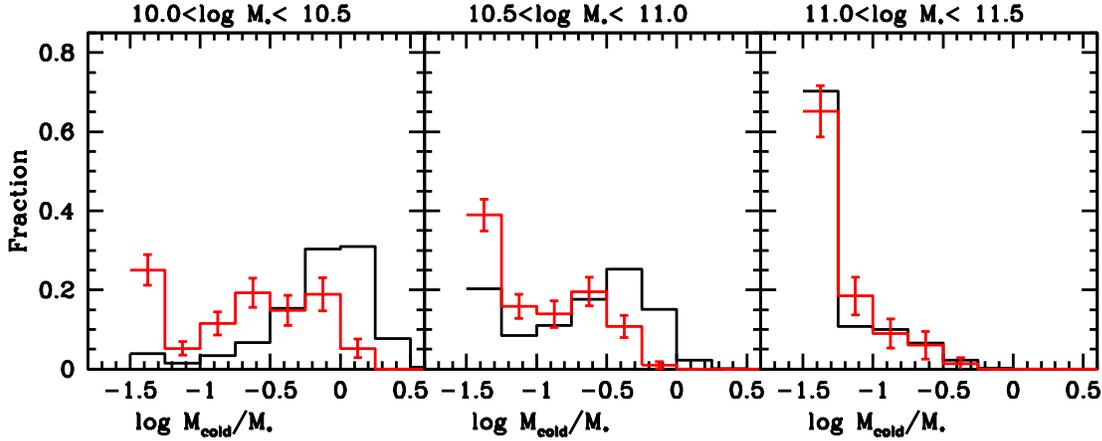}
\caption{ 1. Black histograms show total gas mass fractions evaluated within 
twice the stellar half mass radius for galaxies in the Illustris
simulation. Results are shown in three different stellar mass bins. Red histograms 
show total (atomic+molecular) cold gas fractions in
the same three mass bins, computed from the public release catalogues 
of 350 galaxies from the COLD GASS survey (Saintonge et al
2011). Error bars on the data histograms have been computed using boot-strap resampling. 
The error bars on the Illustris histograms
are not shown because they are negligibly small. 
\label{models}}
\end{figure*}

We work with the ``cutouts'' provided by the Illustris data base, that
include all the dark matter, gas, star, wind and black hole information for
both subhalos and the parent friends-of-friends halo for each galaxy.
Unless indicated otherwise, the plots in section 3 are generated from the
subhalo cutouts. We have excluded galaxies in clusters and massive groups
from our analysis, so in the majority of cases the majority of the
particles in the parent halo are bound to the subhalo of the central
galaxy.

We implement an algorithm to find the two-dimensional
projection of the stellar mass distribution tion with
the least scatter about the major axis and we call this the
best edge-on projection.
We begin by locating the centre-of-mass of the stellar distribution and
translating the positions of all the particles so that the coordinate
(0,0,0) is located at this position. We then proceed to find the best
edge-on projection of the stellar distribution. This is done by projecting
the cube along the x,y and z axes in turn, performing a linear fit to the
stellar mass distribution for each 2-dimensional ``sky'' projection. 
The linear fit with the minimum scatter  defines the x-axis of our new, chosen
projection. This procedure makes sure we start with
a galaxy with a well-defined major axis when viewed in two 
dimensions. The linear fit is carried out using standard
least-squares regression, treating each grid cell above a given
projected stellar mass density threshold as a separate data
point. \footnote {We adopt $\log \mu_* = 10^7 M_{\odot} kpc^{−2}$ as our threshold, but find
that the outcome is insensitive to this exact choice.} 
The stellar distribution is then rotated around the new x-axis,
until the scatter of the stellar distribution is minimized.  This defines
the best edge-on projection for the stars.

We then create two-dimensional binned maps of the projected density of
stars and gas, as well as gas-mass and stellar-mass weighted metallicity
and velocity maps. An example of such a map for a galaxy with gas mass
fraction comparable to that of the Milky Way ($f_g=0.1$) is shown in Figure
2.  The 2D binning in the x and y-directions is done in logarithmic units
with a bin size of 0.15 in $\log x$ and $\log y$. The contribution from gas and stars located at
$x,y<1$ kpc is placed into the central bin, creating the spiky pattern
seen in the stellar surface density map. This spiky feature may look somewhat artificial, but recall
that the spatial resolution of the Illustris-1 simulation is $\sim 1$
kpc, and making projected density plots in logarithmic units
has the advantage of allowing us to visualize a wide range
of spatial scales in a single diagram. 

From top-to-bottom in the left-hand column, we show total gas density, neutral
gas column density and neutral gas-weighted velocity maps. In the
right-hand column, we show two-dimensional maps of  gas-mass weighted
temperature, stellar surface mass density and gas-phase metallicity. 
The neutral hydrogen fractions and gas temperatures
for each cell are read directly from the simulation outputs.
In particular, each gas cell contains the fraction of its hydrogen
mass which is neutral (the NeutralHydrogenAbundance
field), and multiplying this value by the cell mass
times X=0.75 gives an estimate of the total neutral hydrogen
mass. In the Illustris simulation itself, this field is calculated
through an iterative solution to the primordial hydrogen and
helium network of Katz,Weinberg, \& Hernquist (1996) combined
with the metal line radiative cooling processes, AGN
radiative feedback model and UVB heating as described in
Vogelsberger et al. (2013) including the self-shielding prescription
following Rahmati et al. (2013). For dense starforming
gas, this calculation uses the effective temperature
from the equation of state ISM model of Springel \&
Hernquist (2003), which would require further refinement
for detailed analysis in this density regime. No model for
the formation of molecular hydrogen has been included, so
the column density estimates will be over-estimated for the
very central regions with $\log N_H > 10^{22}$ atoms cm${-2}$. This complication
should have negligible impact in the present work as we do
not consider in detail HI in dense disk environments but
instead focus on larger scales.

In the particular galaxy shown in Figure 2,
the stars and the gas are  well aligned,  and
distributed in a plane with both stellar and gas surface densities that
fall off very steeply as a function of perpendicular distance from the disk
beyond a scale heights greater than a few kpc.  Gas with temperatures less
than $10^4$ K is also aligned with the disk extending in patches to scale
heights of 2-3 kpc. The gas-phase metallicity in this galaxy is high: twice
solar in the vicinity of the disk, dropping to a half solar at at a
distance of 10 kpc above the plane. Even though the neutral gas column
density drops steeply at scale heights greater than a few kpc, the bottom
left panel shows that the low density gas is co-rotating with the thin disk
out to scale heights in excess of 10 kpc.

\begin{figure*}
\includegraphics[width=159mm]{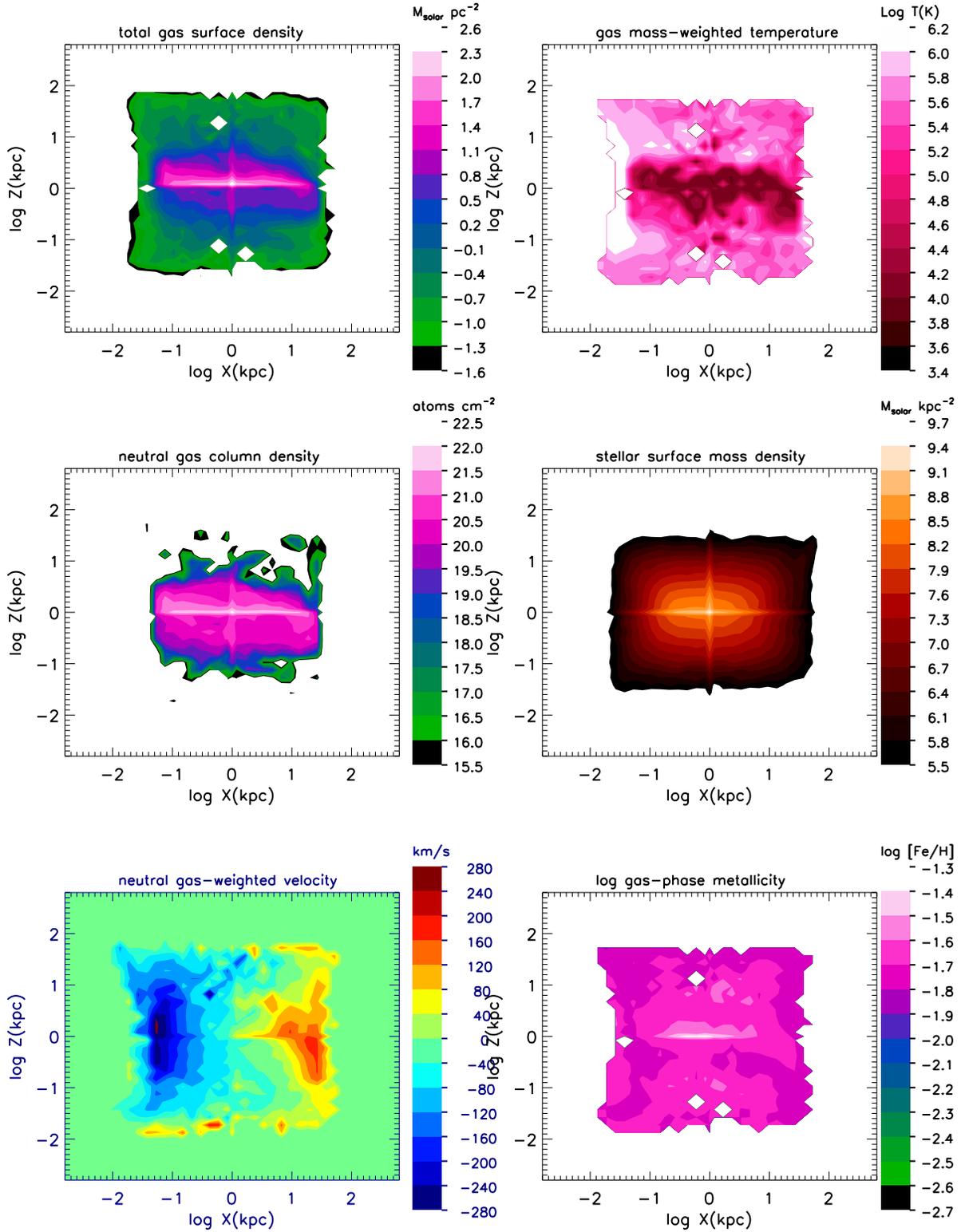}
\caption{ Two-dimensional binned maps of a)projected gas density, b)gas mass-weighted
temperature, c)neutral gas column density, d)stellar surface mass density,
e)neutral gas-weighted velocity, f) gas-phase metallicity. This galaxy has
an ISM gas mass fraction comparable to that of the Milky Way (($f_g=0.1$).
The 2D binning in the x and y-directions is done in logarithmic units
with a bin size of 0.15. The contribution from gas and stars located at
$x,y<1$ kpc is placed into the central bin. 
\label{models}}
\end{figure*}

In the next section, we will define a series of metrics for quantifying the
properties of these 2-dimensional gas maps, and we will study how these
metrics depend on the quantity $f_g$.

\section{Results} \subsection {Radial distributions} We begin by plotting
radial distributions of stellar surface mass density, total gas density,
gas mass weighted temperature, neutral gas density  and gas mass weighted
metallicity is Figures 3-7. In each of these figures, average radial
profiles are plotted in the top left panel. Cyan, blue, green and red
curves show results for galaxies with gas mass fractions $f_g$ in the range
0.3-1, 0.1-0.3, 0.03-0.1 and 0.01-0.03, respectively. In the next three
panels, individual, radially-averaged data points are plotted for each of the 30 galaxies in
each of the gas fraction sub-samples (results are not shown for the lowest
gas fraction sub-sample with $f_g=0.01-0.03$.) These panels allow us to
assess the amount of scatter from one galaxy to another for a given radial bin. 

The samples have been constructed to contain galaxies in narrow range in
stellar mass but with widely different gas mass fraction values evaluated
within twice the optical radius, so it is no surprise that the radial
stellar surface density profiles are the same for all bins in $f_g$ in
Figure 3, but  that the central  gas surface densities scale strongly with
$f_g$ in Figure 4. What is more interesting, is that the differences in
total gas surface density as a function of $f_g$ persist out to radii of
200-300 kpc, i.e. out to the virial radius of the subhalo.  This means that
galaxies with more gas are located in dark matter halos with more gas.
Figure 4 also shows that the galaxy-to-galaxy scatter in total gas surface
density is a factor of two or more larger than the scatter in stellar
surface mass density at all radii.

Figure 5 shows that the average temperature profile of the halo gas depends
strongly on $f_g$.  In halos hosting galaxies with low gas fractions, the
average temperature remains relatively constant at $\sim 10^6$ K all the
way from 10 kpc to 100 kpc. Only at radii less than 10 kpc, does the
average gas temperature drop by a factor of 2-3.  In contrast, in halos
hosting galaxies with high gas mass fractions, the average gas temperature
drops montonically by a factor of 10 from radii of 100 kpc to the center.
It should be noted, however, that there is very large (factor 10-30)
galaxy-to-galaxy scatter in the temperature profiles at all radii and for
all of the $f_g$ sub-samples.

The rise in gas temperature towards the outer regions of halos means that
the neutral hydrogen column density exhibits a stronger radial dependence
than the total gas surface density (6 orders of magnitude drop from R=0 to
100 kpc in $\log N_H$ compared to 4 orders of magnitude drop in total gas
surface density). In addition, the large scatter in gas temperature induces
a corresponding large scatter in $\log N_H$.  Nevertheless, in Figure 6,
clear trends are seen in radial neutral hydrogen column density
distributions as a function of $f_g$ at radial distances less than 100 kpc,
where the fraction galaxies that are surrounded by high neutral column
density gas ($>10^{19}$ atoms cm$^{-2}$) is a strong function of $f_g$. We
will come back to this point in the next section, where we discuss how our
results compare with recent observational findings.

Finally, Figure 7 shows gas phase metallicity trends as a function of
radius for the different $f_g$ sub-samples. The gas-phase metallicity is
lower for galaxies with higher gas mass fractions and the strength of the
effect increases strongly with radius out to the virial radius of the
subhalo.  At R=200 kpc, there is a factor of 3 difference in metallicity
between galaxies with gas fractions comparable to that of the Milky Way and
galaxies in the most gas-rich subsample with $f_g=0.3-1$.

\begin{figure}
\includegraphics[width=91mm]{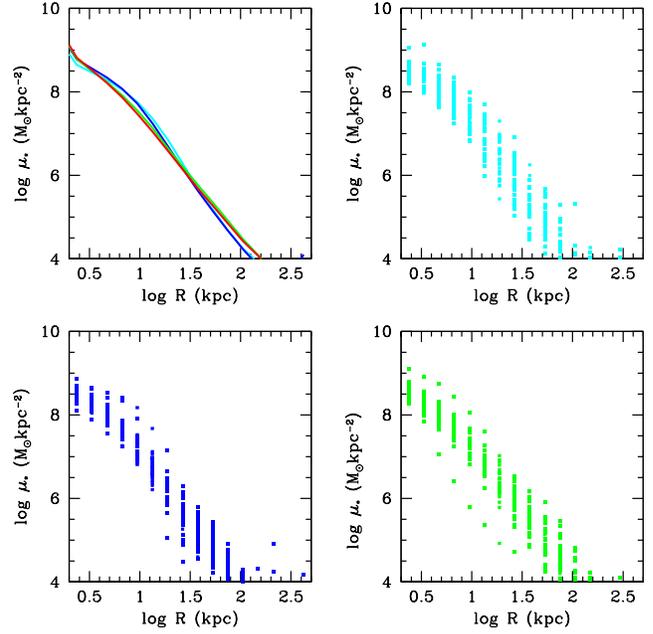}
\caption{ {\em Top left:} Average sellar mass surface density profiles for galaxies
in each of the four subsamples. Results for the subsample with $f_g=0.01-0.03$
are shown in red, $f_g=0.03-0.1$ in green,  $f_g=0.1-0.3$ in blue, and
$f_g=0.3-1.0$ in cyan. In the next three panels,   
individual, radially-averaged data points are plotted for each of the 30 galaxies in
the  $f_g=0.03-0.1$ (green; bottom right),  $f_g=0.1-0.3$  (blue; 
bottom left), and $f_g=0.3-1.0$ (cyan; top right)
\label{models}}
\end{figure}

\begin{figure}
\includegraphics[width=91mm]{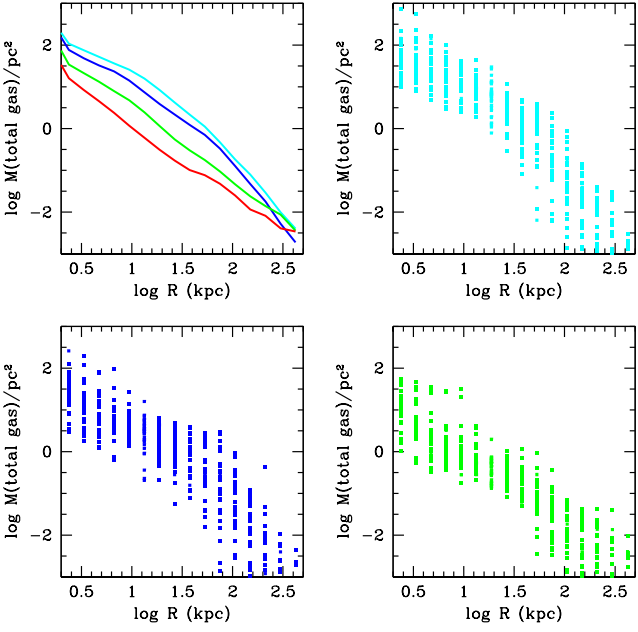}
\caption{ As in Figure 2, except for total gas surface density profiles.  
\label{models}}
\end{figure}

\begin{figure}
\includegraphics[width=91mm]{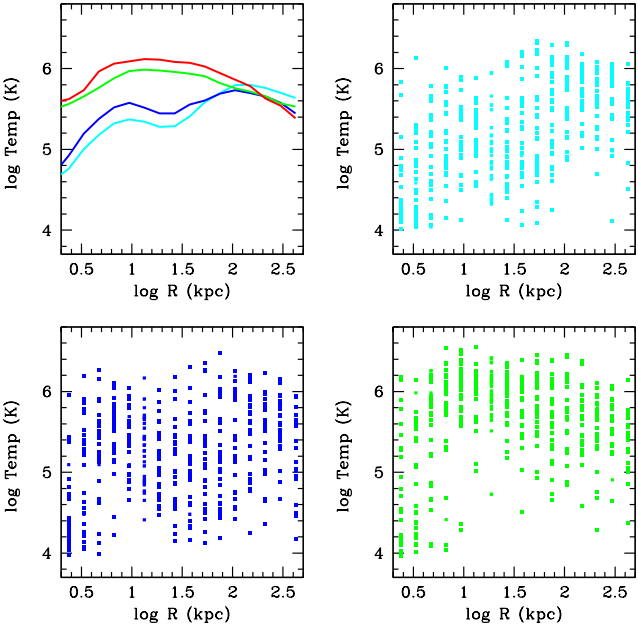}
\caption{ As in Figure 2, except for total gas mass-weighted temperature profiles. 
\label{models}}
\end{figure}

\begin{figure}
\includegraphics[width=91mm]{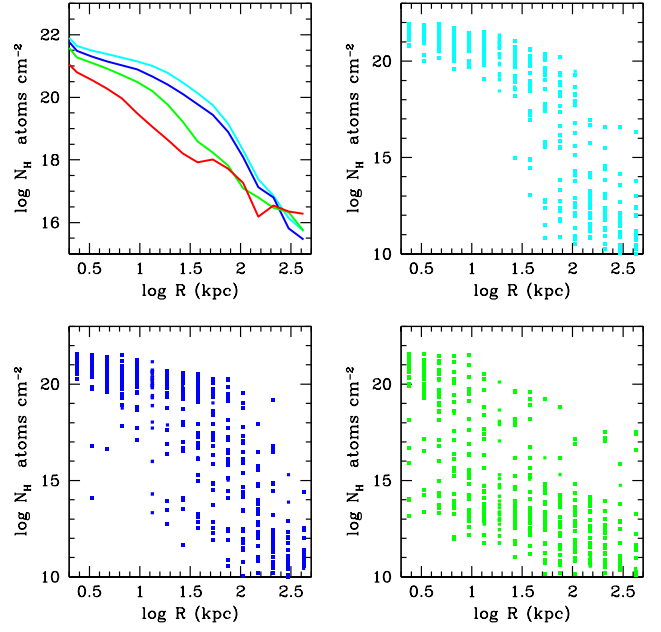}
\caption{ As in Figure 2, except for neutral hydrogen column density profiles. 
\label{models}}
\end{figure}

\begin{figure}
\includegraphics[width=91mm]{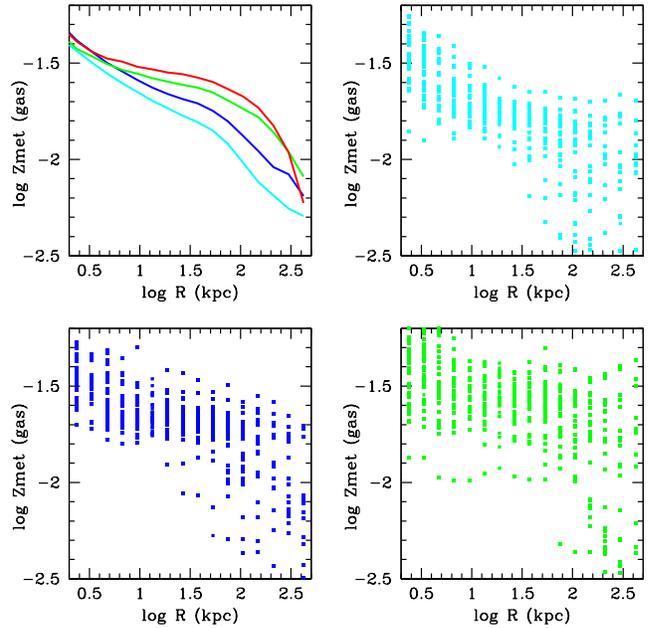}
\caption{ As in Figure 2, except for total gas mass-weighted metallicity profiles. 
\label{models}}
\end{figure}

\subsection {Dependence of radial profiles on orientation} We now explore
the extent to which the radial profiles discussed in the previous
subsection depend on orientation with respect to the stellar major axis of
the galaxy. We compute radial profiles in 6 bins in angle $\phi$: 0-15,
15-30, 30-45, 45-60, 60-75 and 75-90 degrees (colour-coded cyan, blue,
green, black, red, magenta in Figure 8). Results are shown for galaxies in
the two low gas mass fraction subsamples in the left-hand column, and for
the two high gas mass fraction subsamples in the right-hand column. The
results show that gas surface densities show enhancements at angles less
than 30 degrees with respect to the major axis of the disk at radial
distances less than 30 kpc. The effects are significantly stronger in
galaxies with low gas mass fractions compared to galaxies with high gas
mass fractions. Later on we will show that this is because high column
density gas is located in a thickened, rotating structure in these systems.

Gas temperature profiles also show orientation effects. At radial distances
less than 30 kpc, we see the expected effect that gas temperatures are
lower along the disk than perpendicular to the disk. No orientation
dependence in temperature is seen at radii greater than 30 kpc for galaxies
with low gas mass fractions. 
However, an interesting trend is seen for gas
to be hotter in the plane of the disk at radii between 30 and 100 kpc in
the high gas mass fraction galaxies. 
Stellar
winds have preferred outflow directions perpendicular to the
star-forming gas disk, so the result cannot be explained this
way. We hypothesize instead that the higher temperatures
in the plane of the disk may be the
consequence of shock-heating of gas as it falls onto the outer disk. This
shock-heating may also be responsible for the ``double-humped'' form
of the gas temperature profiles seen in both Figure 5 and in Figure 8.  In
follow-up work, we plan to check this hypothesis in detail by following the
trajectories, temperature and star formation histories of gas in the halo
using tracer particles (Genel et al 2013).

\begin{figure}
\includegraphics[width=91mm]{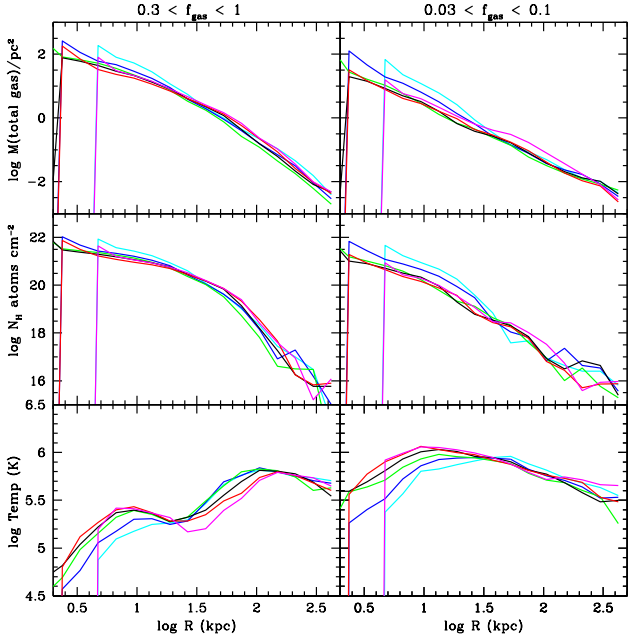}
\caption{ Dependence of radial profiles of total gas surface density (top),
neutral hydrogen column density (middle) and temperature (bottom)
on orientation with respect to the major axis of the disk. Different colour
lines show redial profiles evaluated at different orientation angles:
0-15$^{\circ}$ (cyan), 15-30 $^{\circ}$ (blue), 30-45$^{\circ}$(green),
45-60$^{\circ}$(black), 60-75$^{\circ}$(red), 75-90$^{\circ}$(magenta).     
Results for the  gas-rich subsample with $f_g=0.3-1$ are shown in the
left column and for the gas-poor subsample the  $f_g=0.03-0.1$
in the right column.
\label{models}}
\end{figure}

\subsection {Probes of asymmetry} In this section, we investigate whether
gas-rich and gas-poor galaxies differ in the asymmetry of their
circumgalactic gas distributions. We study asymmetries by comparing gas
distributions above and below the galactic plane, as well as to the right
and to the left of the minor axis of the galaxy in its edge-on projection.

We begin by computing radii along the major axis $R_x(90), R_x(95)$ and
$R_x(99)$ enclosing 90\%, 95\% and 99\% of the in-plane projected stellar
mass, defined as the total stellar mass enclosed within the radii -1 kpc
$<$y$<$ 1 kpc. Material with $x <R_x(90)$ is said to be associated with the
inner disk; material with $R_x(90)<x< R_x(95)$ is said to be associated
with the outer disk , while material with $R_x(95)<x < R_x(99)$ is said to
be associated with the far outer disk .

We define the asymmetry indices  A$_c$ (up-down 90), A$_c$ (up-down 95),
A$_c$ (up-down 99), A$_c$ (right-left 95) and A$_c$ (right-left  99), where
``up-down'' means above and below the galactic place , ``right-left'' means
on either side of the disk minor axis, and 90, 95, 99 refer to inner, outer and
far outer disk, respectively. To compute the indices, we integrate up the
total gas columns from a scale heights of 2kpc above/below  the plane out
to a distance of 200 kpc, and compute (for example) A$_c$ (up-down 90) as $
\log (\left|M_{tot}(up)-M_{tot}(down)\right|/[M_{tot}(up)+M_{tot}(down)])$,
i.e.  the  index is expressed in terms of the logarithm of the fractional difference in
total gas mass above and below the central disk. The other 4 indices are
defined in similar fashion. \footnote {We looked into the effect of changing the 200 kpc upper bound
and found that the main conclusions presented in this section
remain unchanged.}

Figure 9 shows distributions of 4 of these indices for the neutral gas
distribution. Results are shown for gas-rich galaxies with $f_g>0.1$ (blue
histograms) and for gas-poor galaxies with $f_g<0.1$ (red histograms). As
can be seen, the asymmetries are always larger for the gas-poor galaxies.
Half of all gas-poor galaxies have fractional asymmetries above and below
the plane  greater than 0.3, compared to 16\% of the gas-rich galaxies.

In Figure 10, we plot a restricted set of asymmetry indices for a variety of
different quantities, including total gas, gas metallicity and gas
temperature.  As can be seen, the neutral gas has the largest asymmetry
values and the gas metallicity the smallest. Differences between gas-rich
and gas-poor subsamples are only significant for the neutral gas. 
In gas-poor galaxies, the fraction of galaxies with neutral hydrogen
asymmetry parameters close to zero, indicating very
large differences in neutral hydrogen mass on either side of
the disk plane, reaches 50\%. This is not seen for the total
gas asymmtery parameter distribution, nor for any other
quantity. As we
will show in section 3.5 , there is significantly greater injection of
energy into the CGM by radio bubble feedback in the gas-poor subsample.
The injection of bubbles of hot gas introduces irregularities in the
neutral gas distribution, resulting in higher asymmetry indices for that quantity,
but does not greatly affect the total gas distribution.

\begin{figure}
\includegraphics[width=91mm]{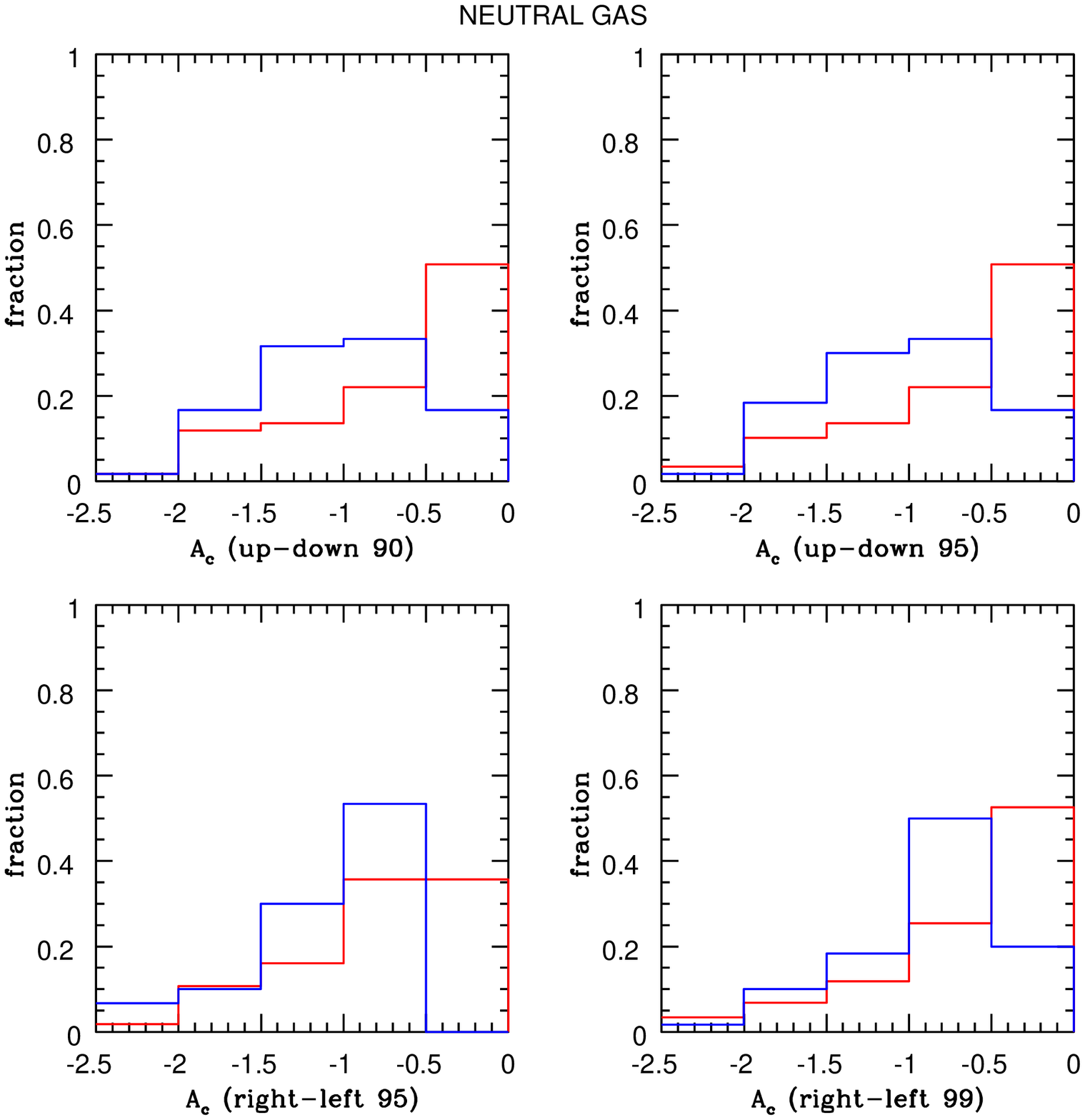}
\caption{ Distributions of 4 different asymmetry indices for the
neutral gas distribution in gas-rich
($f_g>0.1$; blue histograms) and gas-poor ($f_g<0.1$; red histograms) galaxies.
The top left panel shows the asymmetry in the central gas distribution above
and below the plane; the top right panel shows the asymmetry in the outer disk
gas distribution above and below the plane; the bottom left panel shows
the asymmetry in the outer disk to the right and to the left of the disk minor axis;  
the bottom right panel shows the  asymmetry in the far outer disk to 
the right and to the left of the disk minor axis.  
\label{models}}
\end{figure}

\begin{figure*}
\includegraphics[width=169mm]{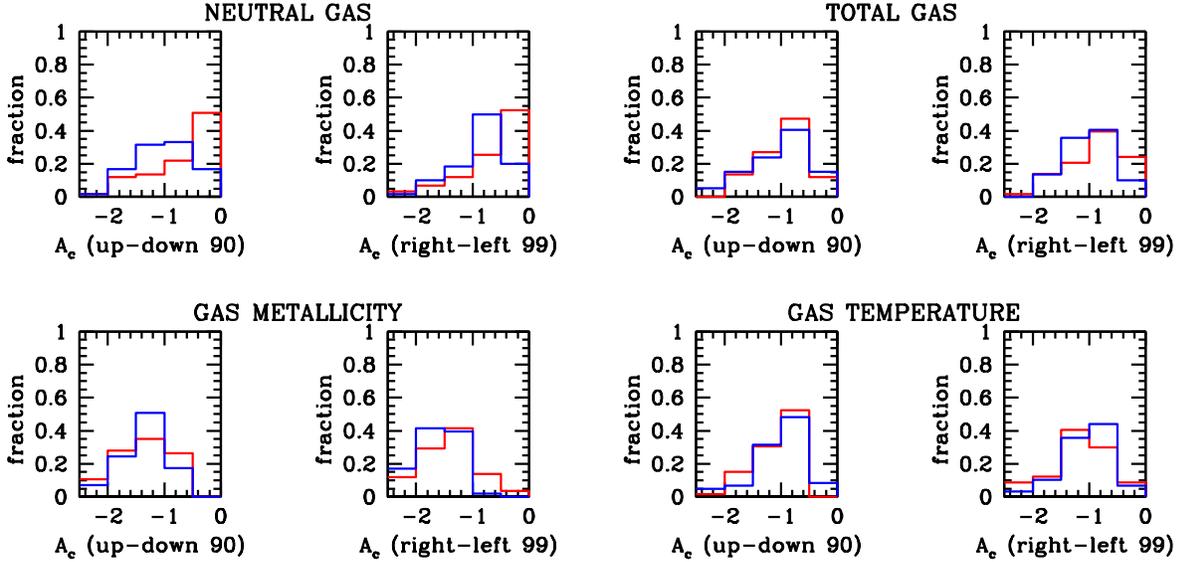}
\caption{ Comparison of  2 different asymmetry indices for the neutral gas
distribution, the total gas distribution, the gas metallicity and the gas temperature.
As in the previous figure, results are shown for gas-rich
($f_g>0.1$; blue histograms) and gas-poor ($f_g<0.1$; red histograms) galaxies.
The two indices represent the 
the asymmetry in the central gas distribution above
and below the plane and 
the asymmetry in the far outer disk to the right and to the left of the disk minor axis.  
\label{models}}
\end{figure*}

\subsection {Kinematic misalignments between gas and
stars and coherent rotation of the CGM} The neutral-hydrogen
weighted velocity maps the galaxies in Illustris show clear signatures of
rotation for all but the most gas poor galaxies when viewed in the edge-on
projection.  Figure 2 shows a case where the velocity minimum and maximum
are well-aligned with the  major axis of the stellar disk, but there are
many cases where the gas disk is clearly misaligned with respect to the
stellar disk.  We quantify these misalignments by locating the velocity
minimum and maximum in the (x,y) plane  and measuring the angles $\phi_1
=arctan \left|y_{min}/x_{min}\right|$ and $\phi_2 =arctan
\left|y_{max}/x_{max}\right|$ with respect to the major axis of the galaxy
viewed edge-on. 

We plot distributions of misalignment angle $\phi$  (measurements
of both $\phi_1$ and $\phi_2$2 are included) in the top right panel
of Figure 11. Cyan, blue, green and red histograms are for
the four sub-samples with different gas mass fractions. The
distribution of misalignment angles is surprisingly broad in
Illustris. Only in galaxies with intermediate gas mass fractions
($f_g$ = 0.03-0.3) is the the gas disk aligned with the
stellar disk to better than 30 degrees in the majority of
cases. The most extreme misaligments are found in the most
gas-poor galaxies. One possible explanation for these results
is that AGN feedback processes are stirring up the gas in
gas-poor objects and accretion is bringing in new gas, misaligned
with the older stellar disk in gas-rich galaxies.

We also measure the full width half maximum (FWHM)
of the vertical coherence length of the rotation by centering
at $x_{min}$ and $x_{max}$ and finding the distance y over which the
velocity is greater than half its minimum/maximum value.
This procedure will underestimate the true coherence length
if the gas is very misaligned with the stars, but it nevertheless
provides a first-order measure than can be compared
between our different sub-samples.

The distribution of FWHM vertical extent of coherent
rotation is shown in the top left panel of Figure 11 and is a
very strong function of the $f_g$: gas-poor galaxies have a median
rotational coherence length of 10 kpc and this increases
by a factor of 4 to 40 kpc in the most gas-rich subsample.
The most extreme objects in the gas-rich subsample have
gas that is coherently rotating over scales of 70-100 kpc! The
distribution of mis-alignment angles is surprisingly broad in
Illustris. Only in galaxies with intermediate gas mass fractions
($f_g = 0.03−0.3$) is the the gas disk aligned with the
stellar disk to better than 30 degrees in the majority of cases.
The most extreme mis-aligments are found in the most gas-poor
galaxies.

In the bottom two panels in Figure 11, we plot the FWHM vertical extent
of coherent rotation as a function of the maximum circular velocity of the
gas.  Gas-poor systems with $f_g < 0.1$ are plotted in the bottom-left
panel and gas-rich systems with $f_g > 0.1$ are plotted in the bottom-right
panel. In the gas-poor galaxies, the coherent extent of the rotation is
generally larger for galaxies with larger $V_{max}$, but no correlation is
apparent for gas-rich galaxies.  This may indicate that the gas has not yet
settled into full rotationally-supported equilibrium in the gas-rich
systems, as might be expected if the gas is in the process of accreting
onto the disk. This will be explored in detail in future work.

\begin{figure}
\includegraphics[width=91mm]{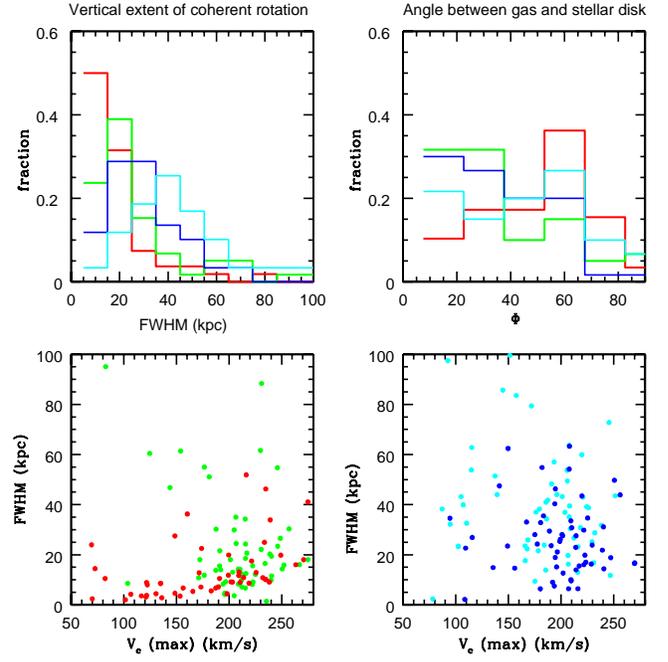}
\caption{ {\em Top left:} Histograms of the distribution of the 
full width half maxmum (FWHM) vertical rotational extent  
of the neutral gas for galaxies in different subsamples
($f_g=0.01-0.03$ in red; $f_g=0.03-0.1$ in green; $f_g=0.1-0.3$ in blue;
$f_g=0.3-1$ in cyan). {\em Top right:} Histograms of the mis-alignment
angle between the gaseous and stellar disks.
In the bottom panels, the FWHM vertical extent
of coherent rotation is plotted as a function of the maximum circular velocity of the
gas for gas-poor galaxies with $f_g < 0.1$ (left) and for gas-rich galaxies with
$f_g > 0.1$ (right).  
\label{models}}
\end{figure}

\subsection{Black holes, radio bubble feedback and gas content} 
In Figure
12, we show the locations of all the Milky Way mass galaxies in our study
in the plane of black hole mass versus black hole accretion rate in
Eddington units. The gas-poor ($f_g < 0.1$) systems are plotted in red and
the gas-rich ($f_g < 0.1$) systems are plotted in blue. One immediately
apparent problem is that the median black hole mass of
the galaxies in our sample is $\sim 10^8 M_{\odot}$, i.e. a factor 24 more
massive than the black hole in our own Galaxy.  
All but 3 of the galaxies in our sample have
black holes that are a factor of 4 or more massive than the black hole in the Milky Way.
\footnote {The three black holes with masses of a few
$\times 10^6 M_{\odot}$ are all part of binary systems}. 

We find that the black holes in Milky Way mass galaxies
in Illustris are all accreting at a few tenths of a percent
of Eddington or less, and are thus in the regime where radio
rather than quasar-mode feedback is occurring. Gas-poor
galaxies have more massive black holes with higher accretion
rates than gas-rich galaxies, suggesting that radio AGN feedback
is controlling the rate of infall of fresh gas onto these systems.

The predicted CGM properties of the gas-poor galaxies
in the Illustris simulation could be called into question if
black hole growth and AGN feedback physics is not implemented 
correctly. 
The gas-rich Milky Way-type galaxies in our study have smaller black holes
with lower accretion rates $10^{-3}-10^{-6} \dot{M}_{Edd}$ and one might
hope that uncertainties the models for black hole growth and AGN
feedback do not compromise our conclusions about the structure and
kinematics of the CGM in such systems.  

In order to check this hypothesis,
we have extracted a set of 15 galaxies in Illustris that have no black
holes, but high gas mass fractions ($f_g>0.3$). In order to assemble such a
sample, we are forced to decrease the cuts in stellar and halo mass  by a
factor of $\sim$3, i.e the majority of the 15 galaxies would not be
included as part of our original selection. Nevertheless, they are useful
for testing the degree to which our conclusions about the radial profiles
and morphologies of neutral gas in very gas-rich galaxies are subject to
uncertainties in the AGN feedback implementation. The top two panels of
Figure 13 show neutral column density and temperature profiles for the
gas-rich galaxies with no black holes (black dashed curves) compared to
results for our two fiducial gas-rich samples. In the bottom two panels
compare results on the vertical coherence length of rotation and the
misalignment angle between gas and stars.  As can be seen, all the results
are fairly similar. The inner cold region of the disk and the vertical
extent of the coherent rotation are somewhat smaller than in our fiducial
samples, consistent with the fact that the systems with no black holes are
found in lower mass halos.

\begin{figure}
\includegraphics[width=89mm]{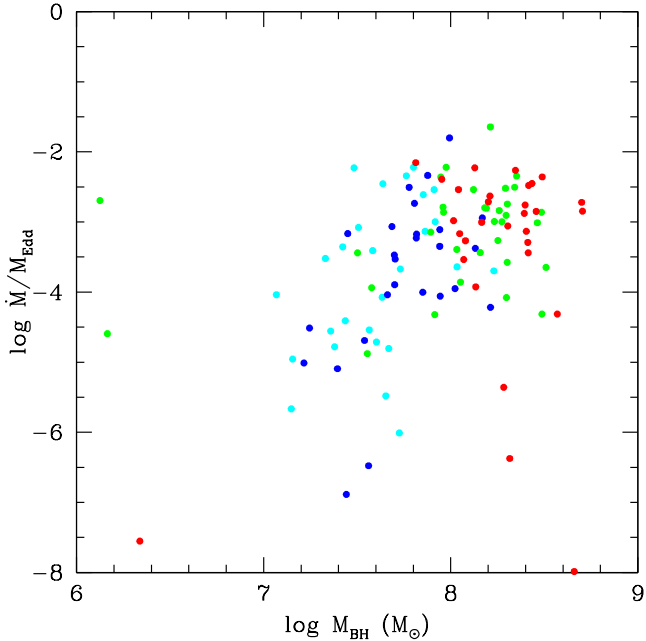}
\caption{ A plot of black hole accretion rate (in Eddington units) as a function
of black hole mass for the supermassive black holes residing in the galaxies
in our sample. Points have been coloured according to the ISM gas mass
fraction of the host galaxy ($f_g=0.01-0.03$ in red; $f_g=0.03-0.1$ in green;
$f_g=0.1-0.3$ in blue; $f_g=0.3-1$ in cyan). 
\label{models}}
\end{figure}

\begin{figure}
\includegraphics[width=92mm]{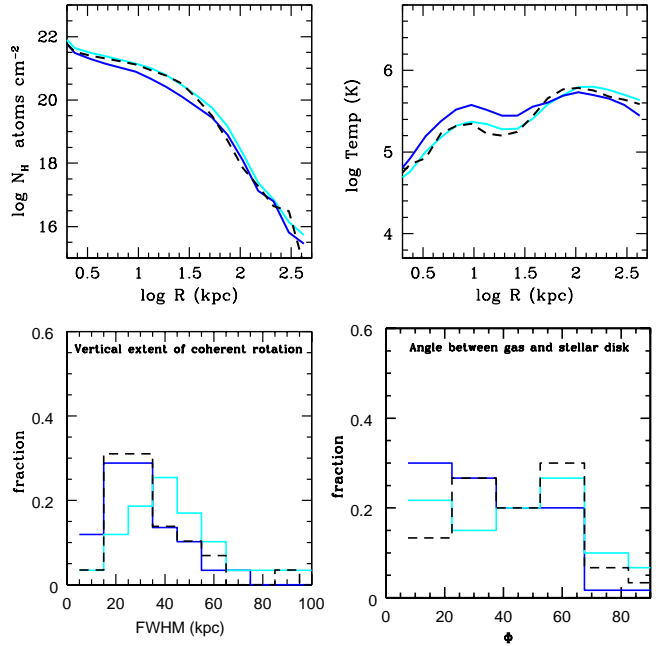}
\caption{ {\em Top panels:} Neutral hydrogen column density and gas temperature profiles
for the fiducial gas rich samples with $f_g=0.1-0.3$ (blue) and $f_g=0.3-1.0$ (cyan)   
and for the gas-rich galaxies with no black holes (dashed black curves).
{\em Bottom panels:} Distributions of the vertical extent of coherent rotation
and the mis-alignment angle between the stellar and gas disks for the same three
samples as in the top panels.
\label{models}}
\end{figure}

\subsection {Summary of simulation results} \begin{itemize} \item The
radial distributions of total gas density, neutral gas density, gas
temperature and metallicity vary strongly as a function of the interstellar medium
gas mass fraction $f_g$.  In gas-rich galaxies with $f_g > 0.1$, 
gas temperatures rise and
gas metallicities drop monotonically from the centre of the
halo out to the virial radius.
Average neutral gas column densities remain
higher than $10^{19}$ atoms cm$^{-2}$ all the way from the center of the
galaxy out to radii of 50-70 kpc, before dropping sharply.  In gas-poor
galaxies with $f_g < 0.1$, gas temperatures remain fixed at $\sim 10^6$ K
and gas metallicities are close to solar all the way from 10 kpc near the
inner disk out to 100 kpc in the outer halo.  The average column density  of
neutral gas begins to drop below $10^{19}$ atoms cm$^{-2}$ at much smaller
radii in gas-poor systems.

\item Gas radial profiles exhibit azimuthal variations at radii less than
30 kpc. These variations are stronger for gas-poor galaxies 
where the gas distribution is more concentrated
along the disk major axis.

\item Gas-poor galaxies have more asymmetric neutral gas distributions than
gas-rich galaxies.

\item  The circumgalactic gas rotates coherently about the center of the
galaxy with a maximum rotational velocity of around 200 km/s.  In gas-rich
galaxies, the average coherence length of the rotating gas is 40 kpc,
compared to 10 kpc in gas-poor galaxies.

\item The gas disk is most often aligned with the stellar disk in galaxies
with intermediate gas fractions of around $\sim 0.1$. 

\item Black hole masses and accretion rates are systematically higher in
gas-poor galaxies than gas-rich galaxies of the same stellar and halo mass.

\end {itemize}

\section {Discussion and comparison with observations}

In this section, we discuss the extent to which our simulation results are
consistent with observational results on the CGM in low redshift galaxies
and present some ideas for future work.

{\em Radial profiles of neutral gas.} The best constraints on the radial
profiles of the neutral gas in the vicinity of nearby galaxies come from
QSO absorption line studies of gaseous halos using the  Cosmic Origins
Spectrograph (COS) on board the Hubble Space Telescope (Tumlinson et al
2013). In general, Ly$\alpha$  absorption is detected for systems with
neutral hydrogen column densities greater than $10^{14}$ atoms cm$^{-2}$. Reasonably
accurate estimates of $N_{HI}$ are possible for damped (DLA) or nearly
damped (subDLA) absorbers with $\log N_{HI}= 18-20$ atoms cm$^{-2}$  by fitting to the
Ly$\alpha$ profiles. It is usually only possible to derive lower limits to
the column density for systems with column densities in the range $ \log
N_{HI}= 14-18$  atoms cm$^{-2}$.  This means that two statistics characterizing the neutral
hydrogen profiles are extractable from the data: 1) the average value of
$N_{HI}$ as a function of radius, since this is mostly set by the highest
column density systems.  2) the so-called {\em covering fraction}, meaning
the fraction of QSO sightlines where a Ly$\alpha$ system with a column
density greater than $10^{14}$  atoms cm$^{-2}$ is detected.

In the top left panel of Figure 14, we plot average neutral hydrogen column
density as a function of radius for simulated galaxies with gas mass
fractions comparable to that of the Milky Way as the solid green curve.
Results for individual galaxies are shown as green points.  We note that
for this plot, $N_{HI}$ is derived from the parent halo cutouts rather than
the subhalo cutouts. This results in higher average column densities at
large radius, because a minority of sightlines do intersect high column
density gas that is not part of the subhalo. The dashed black curve shows
observational results from Werk et al (2014). As can be seen, the radial
run of the average neutral gas column density in the simulations agrees
reasonably well with the observations.

Discrepancies with the data are, however, revealed when we compare the
covering fraction at two different radii. In the middle panel of Figure 14,
green stars indicate covering
fractions of neutral hydrogen with $\log N_{HI} > 14$ atoms cm$^{-2}$ derived from
simulations, while the dashed black line again indicates results from the
COS data. At radii of $\sim 30$ kpc, the covering fractions agree quite
well, but at 100 kpc, the covering fraction of neutral gas in the
simulation is a factor of two below the observations.
We note that Borthakur et al (2015)
compiled statistics for the velocity differences between each
of their Ly$\alpha$ systems and the central velocity of the 21cm
line, and found that almost all systems had $\Delta |V| < 200$
km/s. The maximum velocity difference was 400 km/s, indicating
that the neutral gas is bound within the halo. The
discrepancy between simulations and observations thus cannnot
be explained by the fact that we do not include gas at
very large velocity difference outside the virial radius of the
parent halo.

The COS galaxies span a range of different stellar masses, specific star
formation rates and gas mass fractions, so comparing with Milky Way-type
galaxies in Illustris is not completely accurate. In a recent study,
Borthakur et al (2015) explored the nature and properties of the
circumgalactic medium and its connection to the atomic gas content in the
interstellar medium as traced by the HI 21cm line. A strong correlation
(99.8\% confidence) was found between the gas fraction in the galaxy and
Ly$\alpha$ equivalent width seen in absorption in the associated quasar
spectrum.  In the right panel of figure of Figure 14, we plot the covering
fraction of Ly$\alpha$ absorption systems as a function of R/R(HI), where
R(HI) is the HI radius. In the study of Borthakur et al, the HI radius is
not directly measured, but is inferred from the total HI mass using the
empirical relation of Swaters et al (2002): $\log M_{HI}= 1.86
\log(2R(HI))+6.6$, which has been found to have very small (0.06 dex)
scatter. The data from Figure 9 of Borthakur et al (2015) is plotted in
black and results from the simulation in green.  A strong drop in covering
fraction as a function of R/R(HI) is seen in both observations and
simulations, but the covering factors are a factor of 2 higher in the data
in all but the central bin. The
dashed green line shows what happens if we divide the cold
gas masses in the simulation by a factor of 2 to account for
the fact that the molecular gas can contribute up to half
the total cold gas content of nearby spiral galaxies (Saintonge
et al 2011). As can be seen, this does not alleviate
the mismatch in covering fractions. This shows that the discrepancy
between simulations and data is likely caused by
inaccurate physical prescriptions for gas heating/cooling.

\begin{figure*}
\includegraphics[width=169mm]{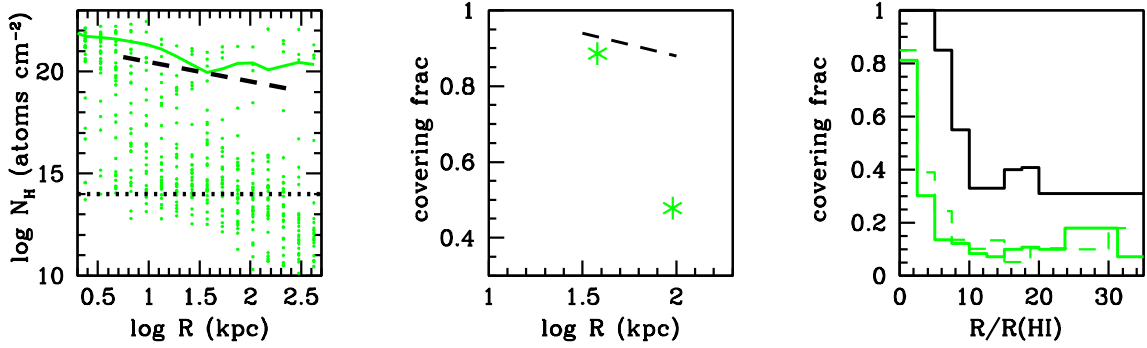}
\caption{ {\em Left:} The solid green curve shows average neutral hydrogen column
density as a function of radius for simulated galaxies with gas mass
fractions comparable to that of the Milky Way.
Results for individual galaxies are shown as green points.  
The dashed black curve shows
observational results from Werk et al (2014). 
{\em Middle:} Green stars indicate covering
fractions of neutral hydrogen with $\log N_{HI} > 14$ derived from
simulations at two diferent radii, while the dashed black line again indicates results from the
COS data.
{\em Right:} The covering
fraction of Ly$\alpha$ absorption systems is plotted as a function of R/R(HI), where
R(HI) is the HI radius (see text for details). The data from Figure 9 of Borthakur et al (2015) is plotted in
black and results from the simulation in green. The dashed green
line shows what happens if we reduce the simulated cold gas masses by a factor 
of two to account for the contribution of molecular gas.
\label{models}}
\end{figure*}

{\em Azimuthal variations in neutral gas density.} Borthakur et al (2015)
also analyzed whether there was any dependence in the properties of the
Ly$\alpha$ absorbers in their sample on the orientation of the sightline
with respect to the major axis of the galaxy, and found no effect. We note
that their sample only includes 7 sightlines that are closer than 30 kpc
from the center of the galaxy. We thus conclude that the simulation results
are in general agreement with the data. 

{\em Extended ionized and neutral gas around edge-on galaxies.} Studies of
edge-on spiral galaxies show that extra-planar material in the form of
diffuse ionized gas (e.g., Rand 1996; Rossa \& Dettmar 2003) and HI (e.g.,
Swaters et al. 1997; Oosterloo et al. 2007) is ubiquitous.  The galaxy NGC
891 is the standard ``poster child'' for extraplanar gas.  Deep HI
observations to a limiting HI column density of $7 \times 10^{19}$ atoms
cm$^{-2}$ (Oosterloo et al 2007) reveal a HI halo containing almost 30\% of
the total neutral gas, which extends to a distance of more than 20 kpc from
the galactic plane on one side of the galaxy.  The halo exhibits regular
differential rotation, at a slightly lower rate than the central disk.
Analysis of optical longslit spectra of the halo of NGC 891 yields further
insight into the physical state of the gas (Rand 1998).  Values of
[OI]/H$\alpha$ indicate that hydrogen is 80-95\% ionized (assuming a gas
temperature of $10^4$ K). Analysis of the vertical dependence of the
[NII]/H$\alpha$, [SII]/H$\alpha$, [OI]/H$\alpha$ and [OIII]/H$\beta$ line ratios
show that simple models where the halo gas is ionized by massive stars in
the disk fail, and some secondary source of ionization must exist. More
recently, two nearby, edge-on spiral galaxies NGC 3044 and NGC 4302 were
observed with the VLA to comparable depths (Zschaechner, Rand \& Walterbos
2015). Rotationally-supported  neutral hydrogen was observed out to a
distance of $\sim$15 kpc in both objects.

We thus conclude that the thick, rotating halos of neutral gas seen in the
Illustris simulations are in general accord with observations of local
edge-on galaxies. However, studies of systematic trends in these halos as a
function of the neutral gas content of the ISM are still lacking. In Figure
15, we show a gallery of 6 neutral hydrogen maps of galaxies with high ISM
gas mass fractions ($f_g=0.3-1$). Neutral gas with column densities in
excess of $10^{19}$ atoms cm$^{-2}$, which would be detectable in deep HI
observations, is coloured white and pink; gas with lower column densities
is coloured blue, green and black. As can be seen, the prediction from
Illustris is for HI halos extending out to 20-30 kpc from the galactic
plane in all such systems, and for significant structure in the HI maps.
Figure 16 shows corresponding neutral gas weighted velocity maps for these
galaxies. There are clear signatures of rotation in all the maps, but the
velocity fields are quite complex in some cases, indicating that the gas in
these systems is likely to be evolving dynamically.  This will be the
subject of future work with gas tracer particles.  Finally, we note that
further modelling of the observed signatures of the ionized gas in
Illustris is also required to make contact with the optical spectroscopic
data.

\begin{figure*}
\includegraphics[width=160mm]{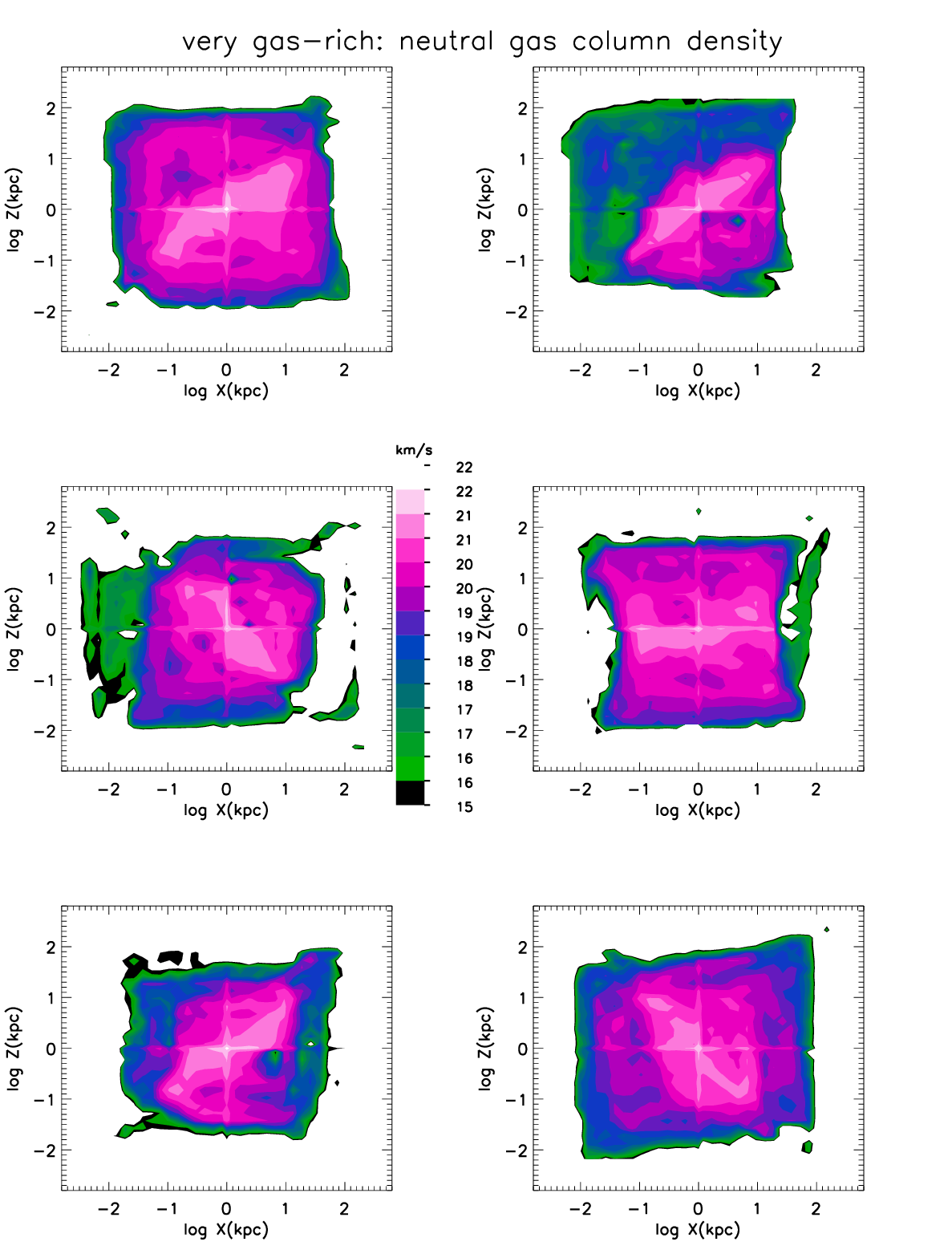}
\caption{ A gallery of two-dimensional binned maps of neutral hydrogen column density
for very gas-rich galaxies with $f_g=0.3-1$.
\label{models}}
\end{figure*}

\begin{figure*}
\includegraphics[width=160mm]{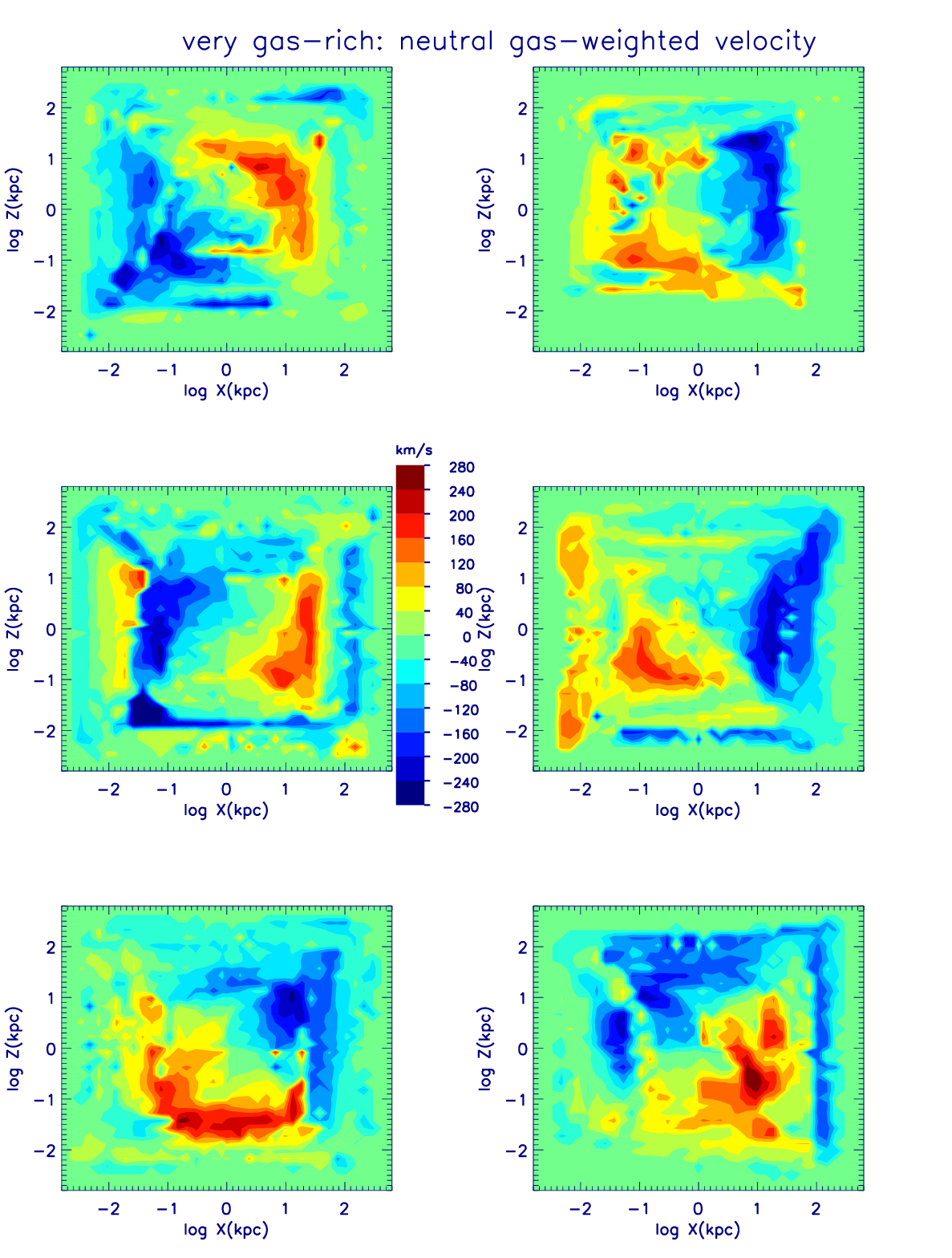}
\caption{ A gallery of two-dimensional binned maps of neutral hydrogen column density-weighted
velocity for very gas-rich galaxies with $f_g=0.3-1$.
\label{models}}
\end{figure*}

{\em Role of heating by radio bubble mode feedback.} We have shown that
gas-poor galaxies in the Illustris simulation have  larger black holes that
are accreting at a higher rate and returning more energy to the surrounding
gas in the form of radio bubbles than gas-rich galaxies of the same stellar
mass. This would imply that the main factor that determines gas accretion
rates onto Milky Way type galaxies at the present day is the energy output
from its supermassive black hole.  Does this paradigm have any
observational support? In 2010, the Fermi Bubbles, clouds of energetic
particles towering 10 kpc above the plane of the Milky Way were discovered
by the Fermi Gamma-Ray Telescope (Su, Slayter \& Finkbeiner 2010). The Fermi Bubbles were
subsequently probed using ultraviolet absorption line spectroscopy by
targeting quasars with sightlines passing through a clear biconical
structure seen in hard X-ray and gamma-ray emission near the base of the
northern Fermi Bubble (Fox et al 2015).  High velocity metallicty
absorption components were detected, supporting a picture where the bubbles
result from a biconical outflow emanating from the Galactic Center.  If
correct, this would imply that the Milky Way's black hole experienced an
outburst 2.5-4 Myr in the past that has resulted in significant heating of
the surrounding gas.  In the Illustris simulations, the radio bubbles are
not launched from the black hole, but are placed at large radii ($\sim$ 100
kpc) from the center of the halo. The greatest heating then occurs close to
the virial radius of the subhalo, which may result in neutral hydrogen
covering fractions at large radii that are too low to match observations.

We conclude, therefore, that there are indications that the Illustris
simulations are producing predictions for the CGM that are in qualitative
agreement with observations, but that significant challenges remain in
matching observations quantitatively. This study has focused on galaxies in
a narrow range in stellar mass, and by extension dark matter halo mass.
Properties of halo gas such as temperature are expected to depend strongly
on mass and this will result in observational signatures such as line
emission changing between low mass and high mass galaxies. Constraints from
large IFU surveys of extra-planar gas in  nearby galaxies (Jones et al, in
preparation) should provide important new constraints in the near future.

Finally, our analysis of the CGM in the vicinity of Milky Way type galaxies
has yielded a picture of a relatively smooth  gaseous halo that is
co-rotating with the central disk.
It may be that when we examine the CGM over a wider range in
both redshift and stellar mass, that we will find physical regimes where
the morphology and kinematics of the CGM is quite different. It will also
be interesting to see how and if the conclusions presented in this paper
depend on  hydrodynamics methodology and implementation of supernova and/or
AGN feedback.

\section*{Acknowledgments}
We thank Chris McKee for helpful discussions.



\begin{thebibliography}{}

\bibitem[\protect\citeauthoryear{Bird et al.}{2014}]{2014MNRAS.445.2313B} 
Bird S., Vogelsberger M., Haehnelt M., Sijacki D., Genel S., Torrey P., 
Springel V., Hernquist L., 2014, MNRAS, 445, 2313 

\bibitem[\protect\citeauthoryear{Borthakur et 
al.}{2015}]{2015ApJ...813...46B} Borthakur S., et al., 2015, ApJ, 813, 46 

\bibitem[\protect\citeauthoryear{Cattaneo et al.}{2007}]{2007MNRAS.377...63C} 
Cattaneo A., et al., 2007, MNRAS, 377, 63 

\bibitem[\protect\citeauthoryear{Faucher-Gigu{\`e}re et 
al.}{2009}]{2009ApJ...703.1416F} Faucher-Gigu{\`e}re C.-A., Lidz A., 
Zaldarriaga M., Hernquist L., 2009, ApJ, 703, 1416 

\bibitem[\protect\citeauthoryear{Faucher-Gigu{\`e}re et 
al.}{2015}]{2015MNRAS.449..987F} Faucher-Gigu{\`e}re C.-A., Hopkins P.~F., 
Kere{\v s} D., Muratov A.~L., Quataert E., Murray N., 2015, MNRAS, 449, 987 

\bibitem[\protect\citeauthoryear{Ford et al.}{2013}]{2013MNRAS.432...89F} 
Ford A.~B., Oppenheimer B.~D., Dav{\'e} R., Katz N., Kollmeier J.~A., 
Weinberg D.~H., 2013, MNRAS, 432, 89 

\bibitem[\protect\citeauthoryear{Fox et al.}{2015}]{2015ApJ...799L...7F} Fox A.~J., et al., 2015, ApJ, 799, L7 

\bibitem[\protect\citeauthoryear{Genel et al.}{2013}]{2013MNRAS.435.1426G} 
Genel S., Vogelsberger M., Nelson D., Sijacki D., Springel V., Hernquist 
L., 2013, MNRAS, 435, 1426 

\bibitem[\protect\citeauthoryear{Guo et al.}{2010}]{2010MNRAS.404.1111G} 
Guo Q., White S., Li C., Boylan-Kolchin M., 2010, MNRAS, 404, 1111 

\bibitem[\protect\citeauthoryear{Hopkins et 
al.}{2014}]{2014MNRAS.445..581H} Hopkins P.~F., Kere{\v s} D., O{\~n}orbe 
J., Faucher-Gigu{\`e}re C.-A., Quataert E., Murray N., Bullock J.~S., 2014, 
MNRAS, 445, 581 

\bibitem[\protect\citeauthoryear{Kamphuis et 
al.}{2013}]{2013MNRAS.434.2069K} Kamphuis P., et al., 2013, MNRAS, 434, 
2069 

\bibitem[\protect\citeauthoryear{Katz, Weinberg, 
\& Hernquist}{1996}]{1996ApJS..105...19K} Katz N., Weinberg D.~H., Hernquist L., 1996, ApJS, 105, 19 

\bibitem[\protect\citeauthoryear{Kere{\v s} et 
al.}{2005}]{2005MNRAS.363....2K} Kere{\v s} D., Katz N., Weinberg D.~H., 
Dav{\'e} R., 2005, MNRAS, 363, 2 

\bibitem[\protect\citeauthoryear{Liang 
\& Chen}{2014}]{2014MNRAS.445.2061L} Liang C.~J., Chen H.-W., 2014, MNRAS, 445, 2061 

\bibitem[\protect\citeauthoryear{Nelson et al.}{2013}]{2013MNRAS.429.3353N} 
Nelson D., Vogelsberger M., Genel S., Sijacki D., Kere{\v s} D., Springel 
V., Hernquist L., 2013, MNRAS, 429, 3353 

\bibitem[\protect\citeauthoryear{Nelson et al.}{2015}]{2015MNRAS.448...59N} 
Nelson D., Genel S., Vogelsberger M., Springel V., Sijacki D., Torrey P., 
Hernquist L., 2015a, MNRAS, 448, 59 

\bibitem[\protect\citeauthoryear{Nelson et al.}{2015}]{2015arXiv150302665N} 
Nelson D., Genel S., Pillepich A., Vogelsberger M., Springel V., Hernquist 
L., 2015b, arXiv, arXiv:1503.02665 

\bibitem[\protect\citeauthoryear{Oosterloo et 
al.}{2007}]{2007A&A...465..787O} Oosterloo T.~A., Morganti R., Sadler E.~M., van der Hulst T., Serra P., 2007, A\&A, 465, 787 

\bibitem[\protect\citeauthoryear{Oppenheimer et al.}{2016}]{2016MNRAS.460.2157O} 
Oppenheimer B.~D., et al., 2016, MNRAS, 460, 2157 


\bibitem[\protect\citeauthoryear{Rahmati et 
al.}{2013}]{2013MNRAS.431.2261R} Rahmati A., Schaye J., Pawlik A.~H., 
Rai{\v c}evic M., 2013, MNRAS, 431, 2261 

\bibitem[\protect\citeauthoryear{Rahmati 
\& Schaye}{2014}]{2014MNRAS.438..529R} Rahmati A., Schaye J., 2014, MNRAS, 438, 529 

\bibitem[\protect\citeauthoryear{Rahmati et al.}{2015}]{2015MNRAS.452.2034R} 
Rahmati A., Schaye J., Bower R.~G., Crain R.~A., Furlong M., Schaller M., Theuns T., 2015, MNRAS, 452, 2034 

\bibitem[\protect\citeauthoryear{Rand}{1996}]{1996ApJ...462..712R} Rand 
R.~J., 1996, ApJ, 462, 712 

\bibitem[\protect\citeauthoryear{Rossa 
\& Dettmar}{2003}]{2003A&A...406..493R} Rossa J., Dettmar R.-J., 2003, A\&A, 406, 493 

\bibitem[\protect\citeauthoryear{Rudie et al.}{2012}]{2012ApJ...750...67R} 
Rudie G.~C., et al., 2012, ApJ, 750, 67 

\bibitem[\protect\citeauthoryear{Saintonge et al.}{2011}]{2011MNRAS.415...32S} 
Saintonge A., et al., 2011, MNRAS, 415, 32 

\bibitem[\protect\citeauthoryear{Schaye et al.}{2015}]{2015MNRAS.446..521S} 
Schaye J., et al., 2015, MNRAS, 446, 521 

\bibitem[\protect\citeauthoryear{Shen et al.}{2013}]{2013ApJ...765...89S} 
Shen S., Madau P., Guedes J., Mayer L., Prochaska J.~X., Wadsley J., 2013, 
ApJ, 765, 89 

\bibitem[\protect\citeauthoryear{Sijacki et 
al.}{2007}]{2007MNRAS.380..877S} Sijacki D., Springel V., Di Matteo T., 
Hernquist L., 2007, MNRAS, 380, 877 

\bibitem[\protect\citeauthoryear{Springel 
\& Hernquist}{2003}]{2003MNRAS.339..289S} Springel V., Hernquist L., 2003, MNRAS, 339, 289 

\bibitem[\protect\citeauthoryear{Springel, Di Matteo, 
\& Hernquist}{2005}]{2005MNRAS.361..776S} Springel V., Di Matteo T., Hernquist L., 2005, MNRAS, 361, 776 

\bibitem[\protect\citeauthoryear{Springel}{2010}]{2010MNRAS.401..791S} 
Springel V., 2010, MNRAS, 401, 791 

\bibitem[\protect\citeauthoryear{Su, Slatyer, 
\& Finkbeiner}{2010}]{2010ApJ...724.1044S} Su M., Slatyer T.~R., Finkbeiner D.~P., 2010, ApJ, 724, 1044 

\bibitem[\protect\citeauthoryear{Suresh et al.}{2015a}]{2015MNRAS.448..895S} 
Suresh J., Bird S., Vogelsberger M., Genel S., Torrey P., Sijacki D., 
Springel V., Hernquist L., 2015, MNRAS, 448, 895 

\bibitem[\protect\citeauthoryear{Suresh et al.}{2015b}]{2015arXiv151100687S} 
Suresh J., Rubin K.~H.~R., Kannan R., Werk J.~K., Hernquist L., 
Vogelsberger M., 2015, arXiv, arXiv:1511.00687 

\bibitem[\protect\citeauthoryear{Swaters, Sancisi, 
\& van der Hulst}{1997}]{1997ApJ...491..140S} Swaters R.~A., Sancisi R., van der Hulst J.~M., 1997, ApJ, 491, 140 

\bibitem[\protect\citeauthoryear{Tumlinson et 
al.}{2013}]{2013ApJ...777...59T} Tumlinson J., et al., 2013, ApJ, 777, 59 

\bibitem[\protect\citeauthoryear{Vogelsberger et 
al.}{2013}]{2013MNRAS.436.3031V} Vogelsberger M., Genel S., Sijacki D., 
Torrey P., Springel V., Hernquist L., 2013, MNRAS, 436, 3031 

\bibitem[\protect\citeauthoryear{Vogelsberger et 
al.}{2014}]{2014MNRAS.444.1518V} Vogelsberger M., et al., 2014, MNRAS, 444, 
1518 

\bibitem[\protect\citeauthoryear{Wang et al.}{2015}]{2015MNRAS.454...83W} 
Wang L., Dutton A.~A., Stinson G.~S., Macci{\`o} A.~V., Penzo C., Kang X., 
Keller B.~W., Wadsley J., 2015, MNRAS, 454, 83 

\bibitem[\protect\citeauthoryear{Weil, Eke, \& Efstathiou}{1998}]{1998MNRAS.300..773W} 
Weil M.~L., Eke V.~R., Efstathiou G., 1998, MNRAS, 300, 773 

\bibitem[\protect\citeauthoryear{Werk et al.}{2014}]{2014ApJ...792....8W} 
Werk J.~K., et al., 2014, ApJ, 792, 8 

\bibitem[\protect\citeauthoryear{Zschaechner, Rand, 
\& Walterbos}{2015}]{2015ApJ...799...61Z} Zschaechner L.~K., Rand R.~J., Walterbos R., 2015, ApJ, 799, 61 

\end{thebibliography}
\end{document}